\RequirePackage{lineno}
\documentclass[superscriptaddress,amsmath,amssymb,nofootinbib]{revtex4}

\usepackage{graphicx}
\usepackage{dcolumn}
\usepackage{bm}
\usepackage{ulem}
\usepackage{xfrac}
\usepackage{enumitem}

\usepackage[printwatermark]{xwatermark}
\usepackage{xcolor}

\usepackage{graphicx,amsmath,amssymb}
\usepackage{epstopdf}
\usepackage{lineno}
\usepackage[colorlinks=true, urlcolor=blue, linkcolor=blue, citecolor=blue]{hyperref}

\usepackage{lipsum}
\usepackage{xfrac}

\begin{document}

\title{Comment on ``Does gravitational confinement sustain flat galactic rotation curves without dark matter?''}
\author{Alexandre Deur}
\affiliation{Department of Physics, University of Virginia,
Charlottesville, Virginia 22901, USA}

\begin{abstract}

We comment on the methods and the conclusion of Ref.~\cite{Barker:2023xsx}, ``Does gravitational confinement sustain flat galactic rotation curves without dark matter?''
The article employs two methods to investigate whether non-perturbative corrections from General Relativity  are important for galactic rotation curves, 
and concludes that they are not. This contradicts a series of articles~\cite{Deur:2009ya, Deur:2016bwq, Deur:2020wlg} that had 
determined that such corrections are large. We comment here that Ref.~\cite{Barker:2023xsx} use approximations known to exclude 
the specific mechanism studied in~\cite{Deur:2009ya, Deur:2016bwq, Deur:2020wlg} and therefore is not testing the finding of Refs.~\cite{Deur:2009ya, Deur:2016bwq, Deur:2020wlg}.

\end{abstract}

\maketitle

\section{Introduction} 
We comment on Ref.~\cite{Barker:2023xsx} which investigates whether corrections stemming from General Relativity (GR)
are important for galactic rotation curves. The work reported in Ref.~\cite{Barker:2023xsx} is motivated by the findings of 
Refs.~\cite{Deur:2009ya, Deur:2016bwq, Deur:2020wlg} that a specific type of non-perturbative mechanism is important for galaxy dynamics. 
That mechanism arises from GR's Field Self-Interaction (FSI). 
It causes the strength of gravity calculated within GR to be larger than the Newtonian expectation.
The effect is found to be important in regimes where the pure Newtonian force is already quite small.\footnote{If this seems conterintuitive, 
consider the example where FSI disrupts gravitational field lines very near a mass $M$, leading to a force $M(1/r+a)/r$. There
is a regime at distance $r$ sufficiently large where $1/r \ll a$ and thus dominates over the Newtonian expectation.} 
FSI
produces, in the case of a disk
galaxy, a logarithmic gravitational potential that directly yields flat galactic rotation curves 
without requiring dark matter (DM). A central aspect of the mechanism emphasized in \cite{Deur:2009ya, Deur:2016bwq, Deur:2020wlg} is its non-perturbative nature. 
Another crucial characteristics of that mechanism is its suppression by the spatial symmetry of the system. 
For example, the amount of missing mass, interpreted as DM in the $\Lambda$CDM model, should be
larger in flatten elliptical galaxies compared to rounder ones. This prediction, made in~\cite{Deur:2009ya}, has now been observed~\cite{Deur:2013baa, Winters:2022ruw}.
The works~\cite{Deur:2009ya, Deur:2016bwq, Deur:2020wlg, Deur:2013baa, Winters:2022ruw} pertain to a more general endeavor that study whether 
observations interpreted as evidence for DM and dark energy (DE)
can instead be jointly explained by the FSI of GR.
The observations studied so far in this context are 
the flat galactic rotation curves~\cite{Deur:2009ya, Deur:2016bwq, Deur:2019kqi, Deur:2020wlg}, 
cluster dynamics~\cite{Deur:2009ya},  
large redshift standard candles~\cite{Deur:2017aas}, 
cosmic microwave background (CMB) data~\cite{ Deur:2022ooc} and 
large structure formations~\cite{Deur:2017aas, Deur:2021ink}.
This framework is denoted  ``GRSI'' in the original articles~\cite{Deur:2022ooc, Sargent:2023djq}, 
but called ``GEFC'' in~\cite{Barker:2023xsx} and we will use this name here. 

To reach the conclusion that FSI is important for galactic rotation curves, Refs.~\cite{Deur:2009ya, Deur:2016bwq} performed static lattice calculations of the GR potential, and Ref.~\cite{Deur:2020wlg} computed that potential 
within  a lensing model based on mean-field technique. The approximations of the former and the modeling of the latter break some of the
tenets of GR. In contrast, the authors of \cite{Barker:2023xsx} strive to be analytical and to preserve GR's basic principles. Consequently, they employ different methods 
from~Refs.~\cite{Deur:2009ya, Deur:2016bwq} and Ref.~\cite{Deur:2020wlg}, whose results they could not reproduce. 
This lead the authors of~\cite{Barker:2023xsx} to refute the validity of~\cite{Deur:2009ya, Deur:2016bwq, Deur:2017aas, Deur:2019kqi, Deur:2020wlg, Deur:2021ink, 
Deur:2022ooc, Sargent:2023djq} and the basic connection of GEFC to GR. 
In what follows, we expand (Section~\ref{Point 1}) on why {\it perturbative} methods like the one used in~\cite{Barker:2023xsx} miss the {\it non-perturbative}  effects of FSI of GR. 
Next, (Section~\ref{Point 2}) we discuss the lensing-based model initially developed in~\cite{Deur:2020wlg} and signals two reasons 
why the calculation  in Ref.~\cite{Barker:2023xsx} miss the FSI effects.  
Then, (Section~\ref{Point 3}) we argue that GEFC is in fact based on GR.  
%
We then summarize and conclude.

\section{The non-perturbative nature of field line collapse \label{Point 1}} 
The FSI of GR discussed in~\cite{Deur:2009ya, Deur:2016bwq, Deur:2017aas} are fundamentally non-perturbative effects, 
and therefore overlooked by the usual perturbative Post-Newtonian (PPN) formalism~\cite{Einstein:1938yz} used in Ref.~\cite{Barker:2023xsx}. 
This was specifically signaled in~\cite{Deur:2016bwq, Deur:2020wlg}, which we quote for convenience: ``{\it A non-perturbative component to the potential would not be identified using the usual \textbf{perturbative} (\textbf{post-Newtonian}) approximation}''~\cite{Deur:2016bwq}. 
And in ~\cite{Deur:2020wlg}: ``{\it Inspecting the \textbf{perturbative post-Newtonian} Lagrangian for two masses $M_1$ and $M_2$ separated 
by $r$ reveals  terms such as $V_{1pn}=G^2M_{1}M_{2}(M_{1}+M_{2})/2r^2$ 
($G$ is the gravitational constant) that are not suppressed at small $v$. These terms can be non-negligible if
$M_1$ and $M_2$ are large enough, but for galaxies, they happen to be generally small. However,  terms such as $V_{1pn}$  
are {\it{perturbative}} corrections, i.e. they omit non-perturbative dynamics, and we will show here that they indeed fail to provide  
the full relativistic dynamics  associated with large masses.}'' 
The assertion that the FSI effects discussed here are non-perturbative comes from our knowledge of Quantum Chromodynamics (QCD).
It possesses  a Lagrangian similar to 
GR's which features, in particular, FSI. The two important differences between GR and QCD are that GR has a tensor field and has very weak coupling, 
while QCD has a vector field and a strong coupling~\cite{Deur:2016tte}. 
Thanks to the first difference, viz the cumulative nature of gravitation, the second difference vanishes for massive enough systems. 
GR field line collapse due to FSI is the mechanism studied in Refs.~\cite{Deur:2009ya, Deur:2016bwq, Deur:2017aas} and
it is firmly established that the analogous collapse of QCD field lines, which leads to static quark confinement, is non-perturbative~\cite{the: Greensite conf.}. 
The parallel between GR and QCD is the main reason for  
asserting the non-perturbative nature of the FSI effects, although it is intuitive that field line collapse is a positive feedback process: 
the denser the field lines in a region, the more it attracts other field lines, which increases the region density.
Another fact supporting the non-perturbative nature is that the author of~\cite{Deur:2009ya, Deur:2016bwq, Deur:2017aas} performed both 
non-perturbative (numerical lattice) 
and perturbative (2$^{\rm nd}$ order PPN) calculations of disk galaxies, with the result that while the non-perturbative lattice calculation displayed a dominant 
FSI effect, the PPN result fell comparatively short by 4 orders of magnitude. 

Ref.~\cite{Barker:2023xsx} employs the PPN formalism and thus cannot produce {\it by definition} the mechanism discussed 
in~\cite{Deur:2009ya, Deur:2016bwq, Deur:2017aas}. Expectedly, \cite{Barker:2023xsx} finds no significant GR corrections.  
That PPN overlooks the effect of interest is the very reason a lattice method was chosen in~\cite{Deur:2009ya, Deur:2016bwq}. 
This is sufficient to rebut the conclusion that the PPN calculation of~\cite{Barker:2023xsx} invalidates the findings of~\cite{Deur:2009ya, Deur:2016bwq}. 
Additionally there is likely a more fundamental reason why any analytical method (including, but not limited to, perturbative techniques) cannot solve the problem at hand,
as well as more specific issues in~\cite{Barker:2023xsx}. We discuss below the more general reason first, and then the more specific issues.\\

The problem of field line collapse is of the most challenging type. 
Solving the analogous but simpler QCD problem amounts to solving the quark confinement problem, 
a Millennium Problem~\cite{Millennium Prize Problems} that remains analytically unsolved despite 50 years of effort by  
the physics community~\cite{the: Greensite conf.}. 
The equivalent problem in GR is more complex as 
it involves tensor fields (rather than vector fields as in QCD) and extended sources (rather than pointlike quarks). Furthermore, QCD is much better understood 
phenomenologically, with possibility of controllable experiments and with vastly more data compared to strong-field GR. 
In fact, there may be a fundamental reason that renders analytical methods, including PPN, unable to predict field line collapse: this type of problem 
is closely related to the Yang-Mills gap problem, which was shown to be fundamentally {\it undecidable}, viz it is a realization of G{\"o}del's incompleteness theorem~\cite{Cubitt:2015xsa}. 
Clearly, there is little hope to solve the analogous  GR problem 
analytically, 
so guidance from the simpler, more familiar, and much better understood QCD is important.

~\\
We discuss now more specific issues in~\cite{Barker:2023xsx}: \\
{\bf 1)} The rotation curve calculations in~\cite{Barker:2023xsx} are performed in the equatorial plane of the galactic disk ($z=0$ in Eqs.~(68)-(76). 
Here and henceforth, all equation numbers refer to Ref.~\cite{Barker:2023xsx})
where FSI effects cancel by symmetry~\cite{Deur:2009ya, Deur:2020wlg}. For a disk of finite thickness symmetric about its equatorial plane ($z=0$), field lines at $z=0$ are equally pulled by the mass/energy at $z>0$ and that at $z<0$, 
resulting in no net effect.\footnote{
In practice, despite the fact that FSI effects cancel in the $z=0$ plan of a perfectly symmetric disk galaxy, one would still expect the rotational speed in $z=0$ to be larger than the Newtonian expectation because of gas friction before the star forms and of few-body interactions with stars at $z \neq 0$. However, this is not included in the approaches discussed here.
} To be sensitive to FSI, calculations must be performed off the equatorial plane where FSI effects do not cancel. \\
{\bf 2)} The PPN calculation of~\cite{Barker:2023xsx} is based on Eq.~(15)
which is the matter part of the action. Why matter-field coupling would be relevant to the (FSI-induced) process of field line collapse  is unclear: 
using only the matter-field coupling overlooks field coupling, the very effect of interest,  and clouds comparisons with pure-field lattice result~\cite{Deur:2009ya, Deur:2016bwq}.
In fact, as noted in~\cite{Deur:2009ya}, considering the pure field case\footnote{
Viz, without matter terms but for the static sources (quarks for QCD, masses for GR) which are not degrees of freedom in the problem.} 
is sufficient, based on QCD experience.
It may be useful to note that while the matter coupling term is relevant to a lensing-based model such as the one discussed in Section~\ref{Point 2}, the subsequent 
scalar field approximation (Eq.~(17)) negates the effect of interest because a scalar gravitational field cannot cause lensing.\footnote{ 
A scalar field can couple only to the trace of the stress-energy tensor, whose trace is zero for massless particles (photon or graviton).}
 Thus, Eq.~(15) is inadequate in both contexts of lattice calculation and lensing model.\\
{\bf 3)} About Eq. (14), the GR Lagrangian expanded in a polynomial form and used as the starting point of the lattice calculations~\cite{Deur:2009ya, Deur:2016bwq}, Ref.~\cite{Barker:2023xsx} writes that Eq.``{\it (14) would require the curious condition on (or off) the background}~''. The ``curious condition'' is then given by Eq. (16). However, Eq.~(14) is the exact GR Lagrangian~\cite{Zee2, Zee1, Salam:1974zw}. 
Thus, either the statement of~\cite{Barker:2023xsx} just quoted is incorrect or Ref.~\cite{Barker:2023xsx} has assumed Eq.~(15)  (matter coupling) which is not used in~\cite{Deur:2009ya, Deur:2016bwq}. Therefore, Eq.~(16) is either incorrect or irrelevant.
Then, Eqs.~(15)-(18) (of which only Eq.~(17) is used in~\cite{Deur:2009ya, Deur:2016bwq})
lead to PPN, which overlooks the non-perturbative effects of interest. \\

~

Finally, we comment here on the statement in~\cite{Barker:2023xsx} that~\cite{Deur:2009ya, Deur:2016bwq} uses the weak-field approximation, 
which may explain the reliance of \cite{Barker:2023xsx} on PPN. This is an important and subtle point. 
As mentioned earlier, the lattice calculations~\cite{Deur:2009ya, Deur:2016bwq} use the GR Lagrangian expanded in a polynomial form, Eq.~(14) in~\cite{Barker:2023xsx}.
The starting point to obtain Eq.~(14) is the GR action in its traditional form, 
$S=\int d^4x \sqrt{-g}g_{\mu \nu}R^{\mu \nu}/(16\pi G)$, with $g_{\mu \nu}$ 
the metric, $g=\det g_{\mu \nu}$, $R^{\mu \nu}$ the Ricci tensor, and $G$ Newton's constant. 
In the textbooks~\cite{Zee2, Zee1} a weak-field approximation is indeed used to develop $S$. 
However, it is only a convenient  intermediate step to re-express the GR Lagrangian in a polynomial 
form: {\it as long as  Eq.~(14) is not truncated, it remains the exact Lagrangian of GR.}
In fact, Eq.~(14) can be derived without weak-field approximation by using the Landau-Ginzburg method, see Fig.~1 of~\cite{Deur:2009ya}. 
On may also find in~\cite{Salam:1974zw} a discussion of Eq.~(14) without mention of the weak-field approximation. 
GEFC needs the  GR Lagrangian in a polynomial form to allow for the methods of strong-QCD (lattice techniques). 
Now, since Eq.~(14) must be truncated for practical use in computer programs, the question is ``can it be truncated without 
altering the solution, and if so, what determines the truncation order?'' At least, the cubic term 
$g\phi\partial_{\mu} \phi \partial^{\mu} \phi$ must be included since the first term (quadratic, $\partial_{\mu} \phi \partial^{\mu} \phi$) has no
FSI. It was observed on the lattice that for 
large masses, the calculated potential changes significantly once the cubic term is added to the free-field (quadratic) term, 
but one can verify that further adding the quartic term $g^2\phi^2\partial_{\mu} \phi \partial^{\mu} \phi$ does not change the result.
Omitting the cubic term but keeping the quartic term also produces collapsed field lines. 
Thus, one FSI term in Eq.~(14) is sufficient. Additional FSI terms do not change the potential: already 
collapsed field lines cannot collapse further.
In other words, this shows that the type of vertex coupling the field (cubic, quartic, quintic $\cdots$ vertices)  is unimportant: in the strong-field regime,
all types lead to field line collapse which results in the same potential.\footnote{
While any type of vertex leads to the same potential in the strong-field regime, One may surmise that the onset of field line collapse, viz
the smallest source mass that triggers a collapse, depends on the type and number of vertices in Eq.~(14).} 
Using a truncated Eq.~(14)  
on a lattice is thus not a weak-field approximation and  is not equivalent to PPN, unless the weak-field limit is explicitly taken.

\section{Lensing calculations \label{Point 2}} 
In Ref.~\cite{Deur:2020wlg}, a lensing model based on a mean-field technique was developed as an alternate approach to the lattice 
method~\cite{Deur:2009ya, Deur:2016bwq}. While the method in~\cite{Deur:2020wlg} is not as directly based on GR as~\cite{Deur:2009ya, Deur:2016bwq}, 
it provides an independent way to study whether FSI can be significant enough to be relevant to galaxy dynamics.
Ref.~\cite{Deur:2020wlg} concluded that it was the case, but \cite{Barker:2023xsx} states that the FSI effects in~\cite{Deur:2020wlg} are 
``{\it overstated by three orders of magnitude''}.  However, we list here two reasons for~\cite{Barker:2023xsx} to miss significant FSI.

{\bf 1)} Ref.~\cite{Barker:2023xsx} computed lensing on rays originating at the disk center. As explained in~\cite{Deur:2020wlg} (Appendix B, last paragraph), this is not a valid choice:
for a ray to remain in the galactic disk until reaching its edge, the angle must be very small but at small radius, this makes the rays to not fully
illuminate the surface through which the flux is computed, thereby undercounting the force. Worst, rays would essentially not 
depart from the equatorial plane where, as explained in the previous Section, FSI cancels due to the 
symmetry of matter around that plane. 
Therefore, large angles exit the galactic disk but small angles miss the bending in the most critical region. 
To avoid this problem,\footnote{
Another advantage of using collimated rays arises from the fact that, as can be seen from a galactic ``spider plot'', 
the speed values have a distribution at a given measured radius, in part because the galactic disk is not infinitely thin. 
Rotation curves display the maximum speed recorded at that radius, 
viz the maximum doppler shift is used to determine the speed at that given radius. 
Thus, it should be the maximally lensed ray that provides the speed reported in rotation curves. However,  identifying that ray is 
difficult in practice since the many rays at different angles required to compute a flux dilute the maximal FSI effect, and we would need to know before hand 
the $z$-dependent location of the small area used to computed the flux. The scheme with the collimated rays also solves this problem. } \cite{Deur:2020wlg} computed how lensing affects collimated rays at small but non-zero height.
This neglects the increase of the $z$ of the rays as they cross the galaxy, but it is a small effect since the rays have relatively small angles and since there is little lensing at large galactic radii due to the falling mass density.
{\bf 2)} A second problem is that to assess that three orders of magnitude are missing, \cite{ Barker:2023xsx} computes the magnitude that would be needed 
for parallel field lines, as in Fig. 3 of~\cite{Deur:2020wlg}. However, Fig. 3 is for illustrative purpose. Quoting \cite{Deur:2020wlg}: ``{\it Fig. 3. For this example, we used densities larger than those typical of galaxies to make the bending of the field lines conspicuous. The bending for actual galaxies densities is small but, as explained next, the ensuing effect is magnified at large distances, making its consequence on galactic rotation sizable at large distances.''} 

~

We conclude this Section by underlining that the lensing approach is only a model, developed to provide a more intuitive and more flexible method than lattice. 
For example, the lattice method cannot study systems more complex than a 2-dimensional axially symmetric disk, while the lensing model can investigate thick, inhomogeneous or warped disks, or the effect of a disk bulge. 
Yet, the lensing model misses important aspects of FSI, e.g., it fails to predict the collapse of field lines between two pointlike sources, the prototypical system for field collapse, because in that model, only masses contribute to bending the field lines, the bending from the energy-momentum of the field between the pointlike sources being ignored.

\section{The GR basis of the  GEFC framework \label{Point 3}} 
Ref.~\cite{Barker:2023xsx} writes that GEFC ``{\it is essentially arbitrary, not necessarily descriptive to GR and inconsistent with the nonlinear, static, vacuum EFEs [Einstein field equations].''}  
The first part of this statement has already been commented upon in Section~\ref{Point 1} where we noted that the starting point of  GEFC  is 
the exact GR Lagrangian in its polynomial form, Eq. (14), which is in turn derived from the more familiar Einstein-Hilbert Lagrangian.  
The statement on inconsistency is based on using PPN, which is incomplete and not used by GEFC. For example, \cite{Barker:2023xsx} tests 
GEFC (or rather their PPN version of it) beyond the static case and notes that GEFC ``{\it does not appear to be healthy, though it may seem so [...] 
under the assumption of staticity.}'' But staticity is the motive of GEFC's scalar field approximation. No insight is gained by testing a framework in domains 
where it is not applicable.

In general, to solve any GR problem one must approximate either GR's basic equations or the problem. Usually both are needed.
To obtain the gravitational force, GEFC uses approximation (lattice, stationary system) or model (lensing). 
Evidently, approximations break some of the principles on which the full theory rests.
 For example, lattice QCD breaks Lorentz invariance despite it being one of the foundation of relativistic quantum field theory. 
 Yet, lattice QCD is the leading approach to QCD in its strong regime. 
The question is whether the approximations preserve the phenomenology of interest. To answer this, GEFC was tested on problems with known analytic solutions, namely 
using only the first term of Eq.~(14) (tested for 1, 2 and 3 spatial dimensions), 
the Lagrangian of the Yukawa force, 
PPN at first order, and 
the $\phi^4$ theory. 
Additionally, the GEFC methodology was applied to QCD. 
In each instance, the known solutions were recovered~\cite{Deur:2016bwq} including for the latter the confining QCD static potential (``Cornell potential''\cite{Eichten:1974af}). 
Ref.~\cite{Barker:2023xsx} only discusses GEFC's recovery of the leading order PPN, 
judging it to be accidental, without  explaining why GEFC successfully obtains the other potentials. 
Recovering these potentials, some highly non-trivial, is unlikely to be accidental.

\section{Summary and conclusion} 
Disproving the non-perturbative, non-analytical results~\cite{Deur:2009ya, Deur:2016bwq, Deur:2017aas, Deur:2020wlg} cannot be done by using the perturbative 
PPN framework as done in Ref.~\cite{Barker:2023xsx}.
That GEFC is unrelated to GR is rebutted by showing that GEFC's approximations preserve the relevant features of its starting point, viz GR. 
This is achieved by applying GEFC's method to known cases, which was done for 
7 distinct potentials (free-field in 1, 2 and 3 dimensions, Yukawa force, leading order PPN, $\phi^4$ theory, and QCD).  
Lets consider the following five facts:
\textbf{(i)} FSI, a feature of GR, provides naturally and without invoking DM and DE a unified explanation of the phenomena otherwise requiring DM and DE 
when analyzed in frameworks where FSI is absent (Newtonian gravity) or cancels (cosmological principle);
\textbf{(ii)} FSI makes predictions of novel phenomena that have been subsequently observed~\cite{Deur:2013baa, Winters:2022ruw};
\textbf{(iii)}  Intriguing parallels exist between GR and QCD, both for the theories and the phenomena they control;\footnote{
Specifically, these parallels are between  
\textbf{(a)} the GR Lagrangian,
\textbf{(b)} the observations interpreted as evidence of  dark matter, 
\textbf{(c)} those for dark energy, 
\textbf{(d)} the cosmic coincidence problem~\cite{cosmic_coinc_pb}, 
\textbf{(e)} the Tully-Fisher relation~\cite{Tully:1977fu}, 
\textbf{(f)} galactic matter density profiles on the GR side, and
\textbf{(A)} the QCD Lagrangian,
\textbf{(B)} the magnitude of hadron masses, 
\textbf{(C \& D)} the confinement of QCD forces in hadrons, 
\textbf{(E)} hadron's Regge trajectories~\cite{the:Regge}, \textbf{(F)} hadronic density profiles on the QCD side, respectively. 
}
\textbf{(iv)} FSI provides an innate framework for observations not explained naturally in $\Lambda$CDM\footnote{ 
{\it Inter alia}, the Tully-Fisher and RAR~\cite{McGaugh:2016leg} correlations, Renzo's rule~\cite{Sancisi:2003xt}, cosmic coincidence~\cite{cosmic_coinc_pb}, Hubble tension~\cite{Abdalla:2022yfr}, dwarf galaxies overcounting~\citep{Klypin:1999uc}, absence of direct detection of dark matter particle and absence of natural candidates within particle physics.};
and \textbf{(v)} solving the equivalent QCD problem of determining the increase of the force magnitude due to FSI has been notoriously difficult 
and its resolution remains a leading problem in physics. 
These facts suggest that, even if a calculation yields too small FSI,
it more likely points to an insufficiency of the method, as the PPN in~\cite{Barker:2023xsx}, rather than a failure of GEFC. In the worst case scenario that approximations used in~\cite{Deur:2009ya, Deur:2016bwq, Deur:2017aas, Deur:2020wlg} 
lead incorrectly to conclude that GR's FSI are significant enough, 
then the facts \textbf{(i-v)} would be pointing to GEFC missing the right mechanism
rather than being wrong. In fact, even if the proposed mechanism (FSI) has been misidentified, it would only put GEFC on the same 
footing as $\Lambda$CDM and alternatives, e.g., MOND~\cite{Milgrom:1983ca}, that are not supported by a verified theory. 
It would not affect GEFC's demonstration that alternatives to DM/DE-based models 
are possible even in the era of precision cosmology: contrary to oft-stated,
high-precision observations, e.g., that of the CMB~\cite{Aghanim:2019ame}, do not establish the existence of DM/DE. Another example is the claim that the Bullet Cluster observation proves DM~\cite{Clowe:2006eq} (this article is titled ``A direct empirical proof of the existence of dark matter''). This is disproved by the fact that the observation is naturally expected by GEFC~\cite{Deur:2009ya}, immaterial to whether or not the FSI mechanism is relevant.

That the numerous parallels between cosmology and hadronic physics are purely fortuitous coincidences is unlikely, especially because of the similar theoretical structure of 
GR and QCD. It is injudicious to ignore these leads only because exact calculations are impossible and approximations can be contested. This is 
especially true in light of the issues presently faced by $\Lambda$CDM  and the ability of GEFC to explain astronomical and cosmological 
observations. As GEFC's claims are outstanding and far-reaching, they must be rigorously scrutinized. This is what Ref.~\cite{Barker:2023xsx} undertook but with a method 
not adapted to the problem. 
The way forward is to test GEFC with numerical non-perturbative methods 
and remember that
50 years of trying to solve the similar, but simpler QCD problem has checked many methods. 

 ~

\noindent{\bf Acknowledgements} 
The work [2-9] was is done in part with the support of the U. S. National Science Foundation award No. 1847771, a Dominion Scholar grant from Old Dominion University and an Ingrassia Family Research Grant at the University of Virginia. The author acknowledges the courteous correspondance with W.E.V Barker, the lead author of~\cite{Barker:2023xsx}.

\end{document}